\documentclass[%
aip,
jcp, 
 sd,%
 amsmath,amssymb,
reprint,%
]{revtex4-1}

\usepackage{graphicx}
\usepackage{dcolumn}
\usepackage{bm}

\usepackage{amsmath}	
\begin{document}

\preprint{AIP/123-QED}

\title[Note: Relaxation time below jamming]
{Note: Relaxation time below jamming}
\author{Harukuni Ikeda}
 \email{hikeda@g.ecc.u-tokyo.ac.jp}
\affiliation{ Graduate School of Arts and Sciences, The University of
Tokyo 153-8902, Japan }

\date{\today}

\maketitle

\newcommand{\diff}[2]{\frac{d#1}{d#2}}
\newcommand{\pdiff}[2]{\frac{\partial #1}{\partial #2}}
\newcommand{\fdiff}[2]{\frac{\delta #1}{\delta #2}}
\newcommand{\bx}{\bm{x}}
\newcommand{\ba}{\bm{a}}
\newcommand{\by}{\bm{y}}
\newcommand{\bY}{\bm{Y}}
\newcommand{\bF}{\bm{F}}
\newcommand{\bn}{\bm{n}}
\newcommand{\be}{\bm{e}}
\newcommand{\new}{\nonumber\\}
\newcommand{\abs}[1]{\left|#1\right|}
\newcommand{\tr}{{\rm Tr}}
\newcommand{\HH}{{\mathcal H}}
\newcommand{\ave}[1]{\left\langle #1 \right\rangle}

Like other critical phenomena, the jamming transition accompanies the
divergence of the relaxation time $\tau$. A recent numerical study of
frictionless spherical particles proves that $\tau$ is inversely
proportional to the lowest non-zero eigenvalue $\lambda_1$ of the
dynamical matrix~\cite{ikeda2019universal}. In this note, we derive the
scaling of $\lambda_1$ below the jamming transition point $\varphi_J$ by
solving the linearized dynamical equation. The resultant critical
exponent agrees with a previous theoretical result for sheared
suspension obtained by applying the virtual work theorem to a simple
shear~\cite{degiuli2015}, highlighting the universality of the
relaxation dynamics below jamming~\cite{ikeda2019universal}.

We consider a system consisting of $N$ frictionless spherical particles
in $d$-dimensions interacting with the following potential:
\begin{align}
 &V = \sum_{i<j}\frac{h_{ij}^2}{2}\theta(-h_{ij}),\
 h_{ij} = \abs{\bx_i-\bx_j}-\frac{\sigma_i+\sigma_j}{2},
\end{align}
where $\bx_i=\{x_i^1,\cdots,x_i^d\}$ and $\sigma_i$ denote the position
and diameter of the $i$-th particle, respectively. We consider a quench
dynamics described by the zero temperature Langevin equation without
inertia:
\begin{align}
&\partial_t \bx_{i} (t) = -\nabla_{i} V.\label{134326_26Dec19}
\end{align}
For $t\gg 1$, one observes an exponential decay $\delta \bx_{i} (t) \sim
\be_i^1 e^{-\lambda_1 t}$ where $\lambda_1$ and $\be_i^1$ respectively
denote the lowest non-zero eigenvalue and eigenvector of the Hessian
$\HH_{ij} \equiv \nabla_{i} \nabla_{j} V$ at the steady-state.  The
energy also shows the exponential decay ${V\sim
\sum_{ij}\delta\bx_i\cdot\HH_{ij}\cdot\delta\bx_j/2 \sim \lambda_1
e^{-2\lambda_1 t}}$. From this equation, it follows
that~\cite{lerner2012toward,hwang2020force}
\begin{align}
 \lambda_{1} = -\lim_{t\to\infty}\frac{\partial_t V}{2V} = \lim_{t\to\infty}
 \frac{1}{N}\sum_{i=1}^N \bF_{i}^2,\label{154724_23Dec19}
\end{align}
where we have defined
\begin{align}
 &\bF_{i} = \sum_{j\neq i} \bn_{ij}f_{ij},\
  f_{ij} = -\frac{h_{ij}\theta(-h_{ij})}{\sqrt{\ave{h^2}}},\new
 &\ave{h^2} = \frac{1}{N}\sum_{i<j}h_{ij}^2\theta(-h_{ij}),\
 \bn_{ij} = \frac{\bx_{i}-\bx_j}{\abs{\bx_i-\bx_j}}.
 \label{154710_23Dec19}
\end{align}

At $\varphi_J$, the model barely satisfies Maxwell's stability
criterion: the number of constraints $N_c$ imposed by the contacts of
constituent particles is $N_c= N_f + 1$, where $N_f$ denotes the number
of degrees of freedom without the global translations and
rotations~\cite{goodrich2012finite}. We define the deficit contact
number as $\delta z \equiv (N_f-N_c)/N$, which vanishes at $\varphi_J$
in the thermodynamic limit. Hereafter we derive the scaling of
$\lambda_1$ as a function of $\delta z$.

Motivated by the numerical
observations~\cite{ikeda2019universal,goodrich2012finite}, we make the
following four assumptions: (i) the exponential decay in the long time
limit does not depend on the initial configuration as long as the
contact number at the steady-state is unchanged, (ii) the lowest
non-zero eigenvalue $\lambda_1$ is isolated and much smaller than the
other non-zero eigenvalues $\lambda_1\ll \lambda_n$, (iii) the
eigenvector of $\lambda_1$, $\be_i^1$, is extended when the system is
isostatic $\delta z=0$, and (iv) the power-law scaling $\lambda_1\sim
\delta z^\beta$ persists up to $\delta z\sim 1/N$.

The assumption (i) allows us to construct an initial configuration by
decompressing the configuration at $\varphi_J$. At $\varphi_J$, the
system satisfies the mechanical equilibrium $\bF_i=0$. Now, to get a
configuration just below jamming, we decompress the system until the
system loses the weakest contact, say $f_{12}$. This breaks the force
balance of $i=1$ and $j=2$ particles:
\begin{align}
\bF_i(0) = (\delta_{i2}-\delta_{i1})\bn_{12} f_{12}.\label{113050_2Jan20}
\end{align}
The typical amplitude of $f_{12}$ can be estimated as follows.  First,
it is known that at $\varphi_J$, the distribution of the contact force
$f_{ij}$ follows the power-law scaling:
\begin{align}
 P(f) \sim f^{\theta},
\end{align}
with $\theta=0.423$~\cite{charbonneau2014fractal} (we neglect the
localized contacts which only gives the sub-leading contribution to the
present argument~\cite{degiuli2015}). Then, following
Refs.~\cite{lerner2012toward,degiuli2015,hwang2020force}, by using the
extreme statistics, one can calculate the typical amplitude of $f_{12}$
as
\begin{align}
 \int_0^{f_{12}}P(f)df \sim \frac{1}{N}\rightarrow f_{12} \sim N^{-\frac{1}{1+\theta}}.\label{201318_27Jul20}
\end{align}
For $t\gg 1$, $\bF_i(t)\propto \dot{\bx}_i(t)$ converges to
the eigenvector of $\lambda_1$, suggesting that only the component
parallel to $\be_i^1$ survives
\begin{align}
 \lim_{t\to\infty}\bF_i(t) \sim \left[\sum_{j}\be_j^1\cdot\bF_j(0)\right]\be_i^1.\label{001021_28Jul20}
\end{align}
Although the above equation seems intuitively obvious, a detailed
investigation of the linearized equation reveals that the assumption
(ii) needs to be used here, see the footnote~\footnote{
Starting from the initial condition, Eq.~(\ref{113050_2Jan20}), we
consider the time evolution of $\bF_i(t)$. For this purpose, we expand
as $\bF_i(t)$ as
\begin{align}
\bF_{i}(t) = \sum_{n=1}c_n(t)\be_{i}^n, 
\end{align}
where $\sum_n$ denotes the summation over the non-zero eigenmode, and
$\be_{i}^n$ denotes the $n$-th eigenvector of $\HH$ normalized so that
${\sum_{i}\be_{i}^n \cdot \be_{i}^m = \delta_{nm}}$.  We calculate
$c_n(t)$ by solving the linearized equation $\delta\dot{\bx}_i=
-(\HH\delta\bx)_i$. After some manipulations, we get
\begin{align}
 c_n(t) 
 = \frac{c_n(0) e^{-\lambda_n t}}{\sqrt{N^{-1}\sum_n\lambda_n^{-1}c_n(0)^2 e^{-2\lambda_n t}}},
 \label{174145_28Dec19}
\end{align}
where the initial condition is given by $c_n(0) = \sum_{i}\be_i^n\cdot
\bF_i(0).$ From the assumption (ii), $\lambda_1\ll \lambda_n$, one can
approximate as $\sum_n\lambda_n^{-1}c_n(0)^2\sim
\lambda_1^{-1}c_1(0)^2$. Substituting this into
Eq.~(\ref{174145_28Dec19}), we get
\begin{align}
 \lim_{t\to\infty}c_n(t) \sim \delta_{n1}\sqrt{N\lambda_1}\sim \delta_{n1}c_n(0),
\end{align}
implying $\lim_{t\to\infty}\bF_i(t) \sim c_1(0)\be_i^1 =
\left[\sum_j\be_j^1\cdot\bF_j(0)\right]\be_i^1$.
}. Substituting Eq.~(\ref{001021_28Jul20}) into
Eq.~(\ref{154724_23Dec19}) and using the normalization condition
$\sum_i(\be_i^1)^2 = 1$, we get
\begin{align}
 \lambda_1 \sim
 \frac{1}{N}\left[\sum_{i}\be_{i}^1\cdot\bF_{i}(0)\right]^2
 = \frac{1}{N}\left[
 (\be_{1}^1-\be_{2}^1)\cdot\bn_{12} f_{12}\right]^2.\label{183409_26Dec19}
\end{align}
The extensiveness of $\be_i^1$ (assumption (iii)) requires
$\abs{\be_{i}^1}\sim N^{-1/2}$, which leads to
\begin{align}
\lambda_1 \sim \frac{1}{N^2}f_{12}^2 \sim N^{-\frac{4+2\theta}{1+\theta}}.\label{184040_26Dec19}
\end{align}
Finally, the assumption (iv) allows us to replace $N^{-1}$ with $\delta
z$, leading to
\begin{align}
 \lambda_1 \sim \delta z^{\beta},\label{175737_26Dec19}
\end{align}
with the critical exponent $\beta = \frac{4+2\theta}{1+\theta} = 3.41$.
This is consistent with a previous result based on the virtual work
theorem for a simple shear~\cite{degiuli2015}. Note that, in some
previous works~\cite{lerner2012toward,hwang2020force}, the authors did
not consider the dynamics of $\bF_i(t)$ and concluded that $\lambda_1
\sim N^{-1}\sum_{i}\bF_i(0)^2\sim N^{-\frac{3+\theta}{1+\theta}}\sim
\delta z^{\beta'}$ with $\beta'=\frac{3+\theta}{1+\theta}=2.41$. This is
a wrong result because the theory fails to take into account the
extensiveness of $\be_i^1$ at $\varphi_J$~\cite{degiuli2015}.

The upper critical dimension of the jamming transition is $d_{\rm
uc}=2$~\cite{goodrich2012finite}. In $d=d_{\rm uc}$, the mean-field
theory asymptotically gives the exact result, but there can still be a
logarithmic correction~\cite{goodrich2014,Kenna2004}:
\begin{align}
 \lambda_1^{2d}\sim \delta z^{\beta}\abs{\log \delta z}^\alpha.\label{141351_21Jul20}
\end{align}
There is currently no theoretical prediction for the value of $\alpha$,
but it can be used as a fitting parameter.

\begin{figure}[t]
\begin{center}
 \includegraphics[width=7cm]{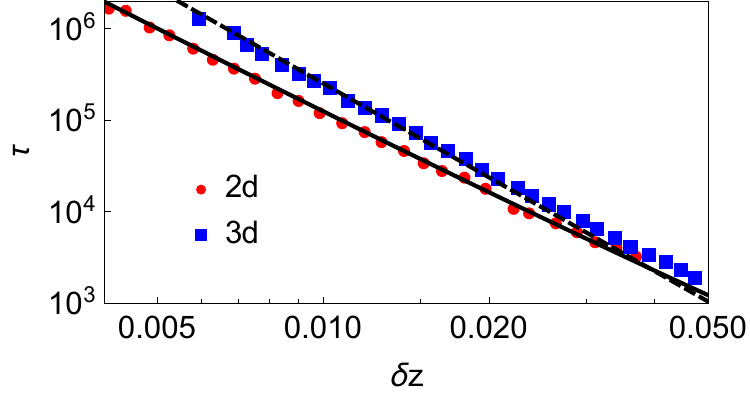} \caption{Scaling of the
 relaxation time $\tau$.  Markers denote numerical results for
 $N=4096$. Solid line denotes $\tau\sim (\lambda_1^{2d})^{-1}\sim\delta
 z^{-3.41}\abs{\log \delta z}^{-2}$, and dashed line denotes $\tau\sim
 \lambda_1^{-1}\sim\delta z^{-3.41}$. Data for numerical results are
 reproduced from Ref.~\cite{nishikawa2020}.}  \label{141325_21Jul20}
\end{center}
\end{figure}

In Fig.~\ref{141325_21Jul20}, we compare our theoretical prediction and
recent numerical results for the relaxation time
$\tau$~\cite{nishikawa2020}, which is inversely proportional to
$\lambda_1$~\footnote{The spatial fluctuation of the contact number
leads to the logarithmic dependence of $\tau$ on
$N$~\cite{nishikawa2020}. It is left as future work to construct a
theory incorporating the spatial fluctuation.}. We find that the
numerical results in $d=3$ are well fitted by
Eq.~(\ref{175737_26Dec19}), while the results in $d=2$ are fitted by
Eq.~(\ref{141351_21Jul20}) with $\alpha=2$.

In summary, we derived the scaling law of the first non-zero eigenvalue
$\lambda_1$, which controls the relaxation time as $\tau\sim
\lambda_1^{-1}$. The result well agrees with the recent numerical result
in $d=3$, while the logarithmic correction is necessary to fit the data
in $d=2$. 

\acknowledgments
We thank A.~Ikeda, Y.~Nishikawa, F.~Zamponi, L.~Berthier, and E.~Lerner
for kind discussions and useful comments. This project has received
funding from the European Research Council (ERC) under the European
Union's Horizon 2020 research and innovation program (grant agreement
n.~723955-GlassUniversality) and JSPS KAKENHI Grant Number JP20J00289.

\subsection*{DATA AVAILABILITY} The data that support the findings of this study are available from the
corresponding author upon reasonable request.

\bibliography{reference}

\end{document}